\documentclass[12pt]{article}
\usepackage{amssymb,amsmath}
\usepackage{epsfig}
\newcommand{\sect}[1]{\setcounter{equation}{0}\section{#1}}

\def\a{\alpha}

\def\rb{\right}
\def\lb{\left}

\newcommand{\eq}[1]{\begin{equation} #1 \end{equation}}

\newcommand{\ml}[1]{\begin{multline} #1 \end{multline}}

\begin{document}

\begin{center}
{\bf{\Large Pulsating strings in warped $AdS_{6}\times S^{4}$ geometry}  \\
\vspace*{.35cm}
}

\vspace*{1cm}
N.P. Bobev${}^{\dagger}$,
H. Dimov${}^{\ddag}$ and
R.C. Rashkov${}^{\dagger}$\footnote{e-mail: rash@phys.uni-sofia.bg}

\ \\
${}^{\dagger}$Department of Physics, Sofia University, 1164 Sofia,
Bulgaria

\ \\

${}^{\ddag}$ Department of Mathematics, University of
Chemical Technology and Metallurgy, 1756 Sofia, Bulgaria

\end{center}

\vspace*{.8cm}

\begin{abstract}
In this paper we consider pulsating strings in warped $AdS_6\times
S^4$ background, which is a vacuum solution of massive type {\bf
IIA} superstring. The case of rotating strings in this background
was considered in hep-th/0402202 and it was found that the results
significantly differs from those considered in $AdS_5\times S^5$.
Motivated by this results we study pulsating strings in the warped spherical
part of the type {\bf IIA} geometry and compare the results with those obtained
in hep-th/0209047, hep-th/0310188 and hep-th/0404012. We conclude
with comments on our solutions and the obtained corrections to the
energy, expanded to the leading order in lambda.
\end{abstract}

\vspace*{.8cm}

\sect{Introduction}

In the last several years the main efforts towards establishing string/gauge theory
duality were focused on $AdS_5\times S^5$. This background has some
particularly interesting features: it is supersymmetric with maximal
amount of supersymmetry, it is dual to $N=4$ SYM theory, the
correspondence between string states and operators on gauge theory
side is particularly well defined, etc. Shortly after the Maldacena
conjecture was established, the supergravity approximation was
intensively studied, but it was realized that the correspondence
requires investigations beyond that limit. The study in  this
direction was developed mainly in  two ways - pp-wave limit of
the background geometry (see for example \cite{bmn}) and semiclassical 
strings in these geometries \cite{gkp}.  The main idea developed in
\cite{gkp} was based on the following. String theory in the most
string backgrounds is highly non-linear and it is impossible by now to
directly attack the problem. One possibility is to consider
semiclassical strings in these backgrounds. In order AdS/CFT
correspondence to be valid, one must consider strings carrying high
energy and large momenta. The simplest way to investigate string/gauge
theory correspondence is to consider an ansatz corresponding
to particular string configurations which makes the problem
tractable. On other side $N=4$ SYM was also intensively studied, for
instance by making use of Bethe ansatz technique
\cite{beis,zar,minzar}.  Comparison of the results on both sides gave good
agreement.

The AdS/CFT correspondence in less supersymmetric backgrounds was
considerably less studied
\cite{rash1,pope,nunhart,moss1,petkou1,schvell}. These
backgrounds however are actually more suitable for description of
hadron physics \cite{sonn1} and therefore are of particular
interest. The main difficulty in working with these backgrounds is
that it is not completely clear which string states to which
gauge theory operators correspond. Nevertheless, it is still interesting to
study the string theory side and to collect results for future
applications. 

In this paper we focus on type {\bf IIA} $AdS_6\times S^4$ background
which contains warp factor. The method of rotating strings 
for $AdS_5\times S^5$ background developed in
\cite{ts}\footnote{For recent review see \cite{tseytlin}.}
was successfully applied to this case in \cite{pope}\footnote{See also \cite{nunhart}.}.
 It was found that
the anomalous dimensions for the operators in the dual D=5 $N=2$ gauge
theory qualitatively differ from those obtained in the case of strings
in $AdS_5\times S^5$. For the short strings
one has the same behavior in both cases
\eq{
E=\sqrt{2J}(1+\frac{1}{8}J + \frac{3}{128}J^2+\frac{1}{1024}J^3+
\cdots).
\label{i1}
}
For long strings in the former case $E$ and $J$ both diverge
logarithmically as $log\epsilon$, while in the latter case $E-J$
approaches a constant
\eq{
E-J=\frac{2}{\pi}-\frac{8}{\pi}e^{-\pi J-2}+\cdots
\label{i2}
}
All these show that for short strings, located near the north pole
the effect of warp factor is negligible and does not affect the
behavior of $E$ and $J$. In the case of long strings, due to the
warp factor, one has repulsion from the equatorial boundary which
restricts the rotation to the $S^3$ part. This is the reason why D=5
corresponding to $AdS_6\times S^4$ Yang-Mills theory is significantly
different from D=4 YM corresponding to $AdS_5\times S^5$.

There is however one more important class of string solutions that are
important and it corresponds to the cases considered in
\cite{min,rashhri,minzar,smedb}, namely pulsating strings. In this
paper we consider pulsating strings in the warped $AdS_6\times
S^4$. In the first Section we review the method of pulsating strings
in $AdS_5\times S^5$ background. In Section 3 we consider
pulsating strings in the warped $S^4$ sphere and find the expression
for the energy $E$ and its quantum corrections. In the Conclusions we
briefly outline the results and comment on some open questions.

\sect{Review of pulsating strings in $AdS_5\times S^5$}

In this section we give a brief review of the pulsating string
solutions obtained first by Minahan \cite{min} and generalized later
in \cite{minzar,rashhri}. For later use we will concentrate only on the case of pulsating
string on the $S^5$ part of $AdS_5\times S^5$, i.e. we consider a circular
string expanding and contracting on $S^5$. We start with the metric of 
$S^5$ and the relevant part of $AdS_5$
\eq{
ds^2=R^2\lb(\cos^2\theta d\Omega_3^2+d\theta^2+\sin^2\theta d\psi^2+
d\rho^2-cosh^2 dt^2\rb),
\label{s.1}
}
where $R^2=2\pi{\a}^\prime\sqrt{\lambda}$. In the simplest
case, we identify the target space time with the worldsheet
one, $t=\tau$, and use the ansatz $\psi=m\sigma$, i.e. the string is
stretched along $\psi$ direction, $\theta=\theta(\tau)$. We will keep the
dependence of $\rho$ on $\tau$ for a while, $\rho=\rho(\tau)$. The reduced Nambu-Goto
action in this case then is
\eq{
S=m\sqrt{\lambda}\int dt \sin\theta \sqrt{cosh^2\rho-\dot\theta^2}.
\label{s.1a}
}
Since we are interesting in calculating the
energy,  it is useful to pass to Hamiltonian
formulation. For this purpose we find first the canonical momenta and 
the Hamiltonian in the form\footnote{For more details see \cite{min}.}
\eq{
H=cosh\rho\sqrt{\Pi_\rho^2+\Pi_\theta^2+m^2\lambda\sin^2\theta}.
\label{s.2}
}
Fixing the string to be at the origin of $AdS_5$ space ($\rho=0$), one
can observe that the squared Hamiltonian have a form very similar to a point
particle. The last term in (\ref{s.2}) can be considered as a
perturbation to the free Hamiltonian, so first we find the wave
function for the free theory in the above geometry
\ml{
\frac{cosh\rho}{sinh^3\rho}\dfrac{d}{d\rho}cosh\rho sinh^3\rho
\dfrac{d}{d\rho}\Psi(\rho,\theta)-\frac{cosh^2\rho}{\sin\theta
  \cos^3\theta} \dfrac{d}{d\theta}\sin\theta\,\cos^3\theta
\dfrac{d}{d\theta} \Psi(\rho,\theta)\\
=E^2\Psi(\rho,\theta).
\label{s.3}
}
The solution to the above equation is
\eq{
\Psi_{2n}(\rho,\theta)=(cosh\rho)^{-2n-4}\, P_{2n}(\cos\theta)
\label{s.4}
}
where $P_{2n}(\cos\theta)$ are spherical harmonics on $S^5$ and the
energy is given by
\eq{
E_{2n}=\Delta=2n+4.
\label{s.5}
}
Since we consider highly excited states, one should
take large $n$, so one can approximate the spherical harmonics by simple
trigonometric functions
\eq{
P_{2n}(\cos\theta)\approx \sqrt{\frac{4}{\pi}}\cos(2n\theta).
\label{s.6}
}
The correction to the energy can be obtained by making use of perturbation
theory, which to first order gives
\eq{
\delta E^2=\int\limits_0^{\pi/2}d\theta\,
\Psi_{2n}^\star(0,\theta)\,m^2\lambda \sin^2\theta\,
\Psi_{2n}(0,\theta)
=\frac{m^2\lambda}{2}.
\label{s.7}
}
Up to first order in $\lambda$ we find for the anomalous dimension of
the corresponding YM operators\footnote{See \cite{min} for more
  details.}
\eq{
\Delta-4=2n[1+\frac 12\,\frac{m^2\lambda}{(2n)^2}].
\label{s.8}
}
It is important to note that in this case the $R$-charge is zero. In order to
take into account the $R$-charge, we consider pulsating string on $S^5$ which has a center
of mass moving on the $S^3$ subspace of $S^5$ \cite{minzar}. While in the
previous example the $S^3$ part of the metric we assumed trivial, now we
consider all the $S^3$ angles to depend on $\tau$ (only). The
corresponding Nambu-Goto action now is
\eq{
S=-m\sqrt{\lambda}\int\,
dt\sin\theta\,\sqrt{1-\dot\theta^2-\cos^2\theta
  g_{ij}\dot\phi^i\dot\phi^j },
\label{s.9}
}
where $\phi_i$ are $S^3$ angles and $g_{ij}$ is the corresponding
$S^3$ metric. The Hamiltonian in this case takes the form \cite{minzar}
\eq{
H=\sqrt{\Pi_\theta^2+\frac{g^{ij}\Pi_i\Pi_j}{\cos^2\theta}
  +m^2\lambda\sin^2\theta}.
\label{s.10}
}
Once again, we see that the squared Hamiltonian looks like the one for a point
particle, however, now the potential has angular
dependence. Denoting the relevant quantum number of $S^3$ and $S^5$ by $J$ and
$L$ correspondingly, one can write the Schrodinger equation for the
free theory 
\eq{
-\frac{4}{\omega}\dfrac{d}{d\omega}\Psi(\omega)
+\frac{J(J+1)}{\omega}\Psi(\omega) = L(L+4) \Psi(\omega),
\label{s.11}
}
where $\omega=\cos^2\theta$. The solution to the Schrodinger
equation is
\eq{
\Psi(\omega)=\frac{\sqrt{2(l+1)}}{(l-j)!}\,\frac{1}{\omega}\left(
  \frac{d}{d\omega}\right)^{l-j}
\omega^{l+j}(1-\omega)^{l-j}
, \quad j=\frac{J}{2}; l=\frac{L}{2}.
\label{s.12}
}
The first order correction to the energy $\delta E$ in this case is
found to be
\eq{
\delta E^2=m^2\lambda\,\frac{2(l+1)^2-(j+1)^2-j^2}{(2l+1)(2l+3)},
\label{s.13}
}
or, up to first order in $\lambda$
\eq{
E^2=L(L+4)+m^2\lambda\frac{L^2-J^2}{2L^2}
\label{s.14}
}
The anomalous dimension then is given by
\eq{
\gamma=\frac{m^2\lambda}{4L}\a(2-\a),
\label{s.15}
}
where $\a=1-J/L$.

We conclude this section referring for more details to \cite{min} and
\cite{minzar}.

\sect{Pulsating strings in warped  $S^4$}

In this section we investigate 
pulsating string solution in warped $AdS_{6}\times S^{4}$ geometry.
It is useful to look at the geometry from different points of
view. From branes point of view this background can be obtained
starting from $N$ $D5$ branes wrapping a circle. After performing
$T$-duality on the circle one ends up with $D4-D8$ system. The
location of the $D8$ branes gives the masses in the hypermultiplet
while the region between two $D8$ branes is described by massive Type
{\bf II A} supergravity \cite{rom1}. The later is also a solution to
the low energy string equations. Ten dimensional background space is
the warped product of $AdS_6$ and $S^4$, i.e. it is a fibration of
$AdS_6$ over $S^4$ and has the isometry group $SO(2,5)\times SO(4)$.

To understand easily the relation to the five dimensional gauge
theories, it is useful to make two steps consistent reduction. The
first step is to integrate over the $S^1$ coordinate yielding
$AdS_6\times S^3$. Next, one can make reduction on $S^3$ gauging its
isometry and ending up with $AdS_6$. On other hand, Romans \cite{rom2}
constructed $AdS_6$ supergravity and the authors of \cite{ferr,oz}
made the connection with $D=5$ superconformal field theory in the
context of AdS/CFT. After these comments let us write the relevant
geometry :
\begin{equation} ds^{2}=\frac{1}{2}W^{2}(\xi)[9(-\cosh^{2}\rho
dt^{2}+d\rho^{2}+\sinh^{2}\rho
d\Omega_{4}^{2})+4(d\xi^{2}+\sin^{2}\xi d\Omega^{2}_{3})]
\label{1.1}
\end{equation}
where
\begin{equation}
d\Omega^{2}_{4}=d\theta^{2}_{1}+\cos^{2}\theta_{1}
(d\theta^{2}_{2}+\cos^{2}\theta_{2}d\phi^{2}_{1}+
\sin^{2}\theta_{2}d\phi^{2}_{2})
\label{1.2}
\end{equation}
\begin{equation}
d\Omega^{2}_{3}=d\theta^{2}+\cos^{2}\theta
d\psi^{2}_{1}+\sin^{2}\theta d\psi^{2}_{2}
\label{1.3}
\end{equation}
are the line elements on the unit $S^{4}$ and $S^{3}$
respectively. The non-trivial fields, i.e. RR four-form $F_{(4)}$ and
the dilaton field supporting this solution, are:
\begin{equation}
F_{(4)}=\frac{20\sqrt{2}}{3}(\cos\xi)^{\frac{1}{3}}\sin^{3}\xi
d\xi\wedge\Omega_{(3)} \,\,\,\, and \,\,\,\,
\exp{\Phi}=(\cos\xi)^{-\frac{5}{6}}.
\label{1.4}
\end{equation}
The explicit form of the warp factor is given by
\begin{equation}
W(\xi)=(\cos\xi)^{-\frac{1}{6}}
\label{1.5}
\end{equation}
The warp factor depends on the above mentioned $S^1$ coordinate $\xi$
and we should note also that in string frame there are no singularities in
the metric. The AdS/CFT correspondence in this background beyond the
supergravity approximation was further studied in \cite{pope}. The
authors of \cite{pope} studied the rotating strings and pp-wave limit
of this geometry. We proceed with the investigation of the class of
pulsating strings in the warped background. The string configuration we
will study is given by the following ansatz
\begin{equation}
t=\tau \qquad \theta=m\sigma \qquad
\rho=\rho(\tau),
\label{1.6}
\end{equation}
i.e. the string is spanned along $\theta$ direction and the time
dependence of $\rho$ realizes the pulsation of the string.
Following the standard procedure developed in \cite{min} 
we write the Nambu-Gotto action for the above metric (\ref{1.1}) 
and string configuration (\ref{1.6})
\begin{equation}
S=-m\sqrt{\lambda}\int dt\frac{\sin\xi}{(\cos\xi)^{\frac{1}{3}}}
\sqrt{9\cosh^{2}\rho-\dot{\rho}^{2}-4\dot{\xi}^{2}}
\label{1.7}
\end{equation}
The canonical momenta that follow from this action are correspondingly
\begin{equation}
\Pi_{\rho}=\frac{m\sqrt{\lambda}\sin\xi}
{(\cos\xi)^{\frac{1}{3}}}\frac{\dot{\rho}}
{\sqrt{9\cosh^{2}\rho-\dot{\rho}^{2}-4\dot{\xi}^{2}}} \label{1.8}
\end{equation}
\begin{equation}
\Pi_{\xi}=\frac{m\sqrt{\lambda}
\sin\xi}{(\cos\xi)^{\frac{1}{3}}}\frac{4\dot{\xi}}
{\sqrt{9\cosh^{2}\rho-\dot{\rho}^{2}-4\dot{\xi}^{2}}}
\label{1.9}
\end{equation}
After some calculations we get for the Hamiltonian the expression
\begin{equation}
H=3\cosh\rho\sqrt{\Pi_{\rho}^{2}+\frac{\Pi_{\xi}^{2}}{4}+\frac{m^{2}\lambda
\sin^{2}\xi}{(\cos\xi)^{2/3}}}
\label{1.10}
\end{equation}
As in the previous section one can observe that $H^{2}$ looks like 
a point-particle Hamiltonian in which the last term serves as a
potential. We will proceed as follows. First of all we will find the
wave function for the free theory and the corresponding energy. After
that we will quantize semiclassically the theory in order to obtain
the corrections to the energy.
We will consider the solutions around $\rho=0$ but
first of all we will find the eigenfunctions and eigenvalues of
the wave equation corresponding to the warped $AdS_{6}\times S^{4}$
geometry.

\subsection{The wave function and corrections to the energy}

Now we are going to derive the wave function corresponding to the free
 theory. The equation in ten dimensions that the wave function should
 satisfy is
\begin{equation}
\triangle_{10}F=\frac{1}{\sqrt{G}}\partial_{\mu}(\sqrt{G}G^{\mu
\nu}\partial_{\nu})F=0
\label{1.14}
\end{equation}
where $\triangle_{10}$ is the Laplace-Beltrami operator for our
metric and $\sqrt{G}$ is 
\begin{equation}
\sqrt{G}=\frac{9^{3}}{2}W^{10}(\xi)\sin^{3}\xi \cos\theta
\sin\theta \cosh\rho \sinh^{4}\rho
\cos^{3}\theta_{1}\cos\theta_{2}\sin\theta_{2}.
\label{1.13}
\end{equation}
For further convenience let us write $\triangle_{10}$ in a more explicit way:
\begin{equation}
\triangle_{10}F=G^{00}\partial^{2}_{0}F+\frac{1}
{\sqrt{G}}\partial_{A}(\sqrt{G}G^{AB}\partial_{B})F+
\frac{1}{\sqrt{G}}\partial_{i}(\sqrt{G}G^{ij}\partial_{j})F
\label{1.15}
\end{equation}
Here we have denoted the spatial part of $AdS_{6}$ indices by
$A,B$, the indices on $S^{4}$ by $i,j$ and the time by $0$. We are
considering motion only on the $S^{4}$ part of the geometry so 
the only non-trivial dependence will be on the coordinates of $S^4$
and on the time $t$. 
Dropping the trivial dependence on the transverse to $S^4$ coordinates,
one can write the equation satisfied by the wave function as
\begin{equation}
G^{00}\partial^{2}_{0}F+\frac{1}
{\sqrt{G}}\partial_{i}(\sqrt{G}G^{ij}\partial_{j})F=0
\label{1.16}
\end{equation}
Explicitly, the equation we have to solve for the free theory is
\begin{equation}
\partial^{2}_{\xi}F+(\frac{4}{3}\tan\xi+3\cot\xi)
\partial_{\xi}F+\frac{1}{\sin^{2}\xi}\triangle_{3}F+\frac{4E^{2}}{9}F=0
\label{1.17}
\end{equation}
Here $\triangle_{3}$ is the Laplace-Beltrami operator for the
"usual" $S^{3}$
\begin{equation}
\triangle_{3}=\frac{1}{\sin\theta
\cos\theta}\partial_{\theta}(\sin\theta
\cos\theta\partial_{\theta})+\frac{1}{\cos^{2}\theta}
\partial^{2}_{\psi_{1}}+\frac{1}{\sin^{2}\theta}
\partial^{2}_{\psi_{2}}
\label{1.18}
\end{equation}
with
\begin{equation}
\triangle_{3}\Psi(\theta)=-s(s+2)\Psi(\theta).
\label{1.19}
\end{equation}
Now we can separate the variables in our equation (\ref{1.18}) by
making the ansatz $F(\xi,\theta)=\Phi(\xi)\Psi(\theta)$ and plug in 
the explicit form of the conformal factor and its derivative:
\begin{equation}
W(\xi)=(\cos\xi)^{-\frac{1}{6}}\,\,\,\,\,\,\,\,\,\,
\frac{dW}{d\xi}=\frac{1}{6}\sin\xi(\cos\xi)^{-\frac{7}{6}}.
\label{1.20}
\end{equation}
The resulting equation is:
\begin{equation}
\frac{d^{2}\Phi(\xi)}{d \xi^{2}}+(\frac{4+5\cos^{2}\xi}{3\sin\xi
\cos\xi})\frac{d\Phi(\xi)}{d \xi}
-(\frac{s(s+2)}{\sin^{2}\xi}-\frac{4}{9}E^{2})\Phi(\xi)=0
\label{1.21}
\end{equation}
where $s$ is the angular momentum on $S_{3}$ and $E$ is the
energy. If we introduce a new variable $x=\tan^{2}\xi$ one can
rewrite the equation (\ref{1.21}) as
\begin{equation}
4x(1+x)\frac{d^{2}\Phi(x)}{dx^2}+\frac{2}{3}(12+13x)
\frac{d\Phi(x)}{dx}-[\frac{s(s+2)}{x}-
\frac{4E^2}{9(1+x)}]\Phi(x)=0.
\label{1.22}
\end{equation}
The solution to this equation is a linear combination of
hypergeometric functions
\ml{
\Phi(x)=C_{1}(1+x)^{\frac{5}{12}-
\frac{\sqrt{25+16E^{2}}}{12}}x^{\frac{s}{2}}\times \\
_{2}F_{1}(\frac{s}{2}
+\frac{5}{12}-\frac{\sqrt{25+16E^2}}{12},\frac{s}{2}+
\frac{19}{12}-\frac{\sqrt{25+16E^2}}{12};s+2;-x)\\
+ C_{2}(1+x)^{\frac{5}{12}-\frac{\sqrt{25+16E^{2}}}{12}}
x^{-\frac{s}{2}-1}\times \\
_{2}F_{1}(\frac{7}{12}-\frac{s}{2}-
\frac{\sqrt{25+16E^2}}{12},-\frac{7}{12}-
\frac{s}{2}-\frac{\sqrt{25+16E^2}}{12};-s;-x)
\label{1.23}
}
We should note that the second term in the
solution should be dropped out because of $-s-2<0$
\footnote{
One can compare (\ref{1.23}) with the expression for the wave function in the case of
"ordinary" $S^{4}$:
\begin{equation}
\Psi^{J}_{L}(\xi)=\tan^{L}\xi \cos^{J}\xi
\,_{2}F_{1}[\frac{L-J}{2},\frac{L-J+1}{2};L+2;-\tan^{2}\xi]\notag
%\label{1.24}
\end{equation}
where $L$ and $J$ are the angular momenta on $S^{3}$ and$S^{4}$
respectively.
}.

Therefore, if we introduce
$J=\frac{\sqrt{25+16E^{2}}}{6}-\frac{19}{6}$ and $L=s$ one can
rewrite $\Phi(\xi)$ as:
\begin{equation}
\Phi(\xi)=C_{1}x^{\frac{L}{2}}(1+x)^{-\frac{J}{2}-\frac{7}{6}}
\,_{2}F_{1}(\frac{L-J}{2},\frac{L-J-\frac{7}{3}}{2};L+2;-x)
\label{1.25}
\end{equation}
with square integrability conditions on
$\Phi(\xi)$
\begin{equation}
J-L=k; \,\,\,\,  k=0,1,2 \cdots
\label{1.26}
\end{equation}
The integrability conditions (\ref{1.26}) mean that the wave function
can be expressed in terms of Jacobi polynomials by making use of the
relation: 
\begin{equation}
P^{(\alpha,\beta)}_{n}(x)=\frac{(n+\alpha)!}
{(\alpha)!(n)!}(\frac{1+x}{2})^{n}
\,_{2}F_{1}(-n,-n-\beta;\alpha+1;\frac{x-1}{x+1}) \label{1.27}
\end{equation}
It is obvious that in our case $n=\frac{J-L}{2}=\frac{k}{2},
\quad  \alpha=L+1, \quad
\beta=\frac{7}{6}$. It is useful also to change the variable in our
solution, by introducing:
\begin{equation}
x=\frac{1-y}{1+y} \,\,\,\, \rightarrow \,\,\,\, 1+x=\frac{2}{1+y}
\label{1.28}
\end{equation}
and rewrite  $\Phi$ in the new variable as:
\begin{equation}
\Phi(y)=C_{1}(\frac{1-y}{2})^{\frac{\alpha-1}{2}}
(\frac{1+y}{2})^{\frac{7}{6}}(\frac{1+y}{2})^{n}
\,_{2}F_{1}[-n,-n-\beta;\alpha+1;\frac{y-1}{y+1}],
\label{1.29}
\end{equation}
or:
\begin{equation}
\Phi(y)_n=\frac{C_{1}(n)!(\alpha)!}{(n+\alpha)!}
(\frac{1-y}{2})^{\frac{\alpha-1}{2}}
(\frac{1+y}{2})^{\beta}P^{(\alpha,\beta)}_{n}(y)
\label{1.30}
\end{equation}
The normalized wave functions $\Phi_{n}(y)$ should satisfy
orthogonality conditions
\begin{equation}
\int\frac{1}{W^{2}(y)}\Phi_{n}(y)\Phi_{m}(y)d\mu(y)=\delta_{nm}.
\label{1.31}
\end{equation}
In (\ref{1.31}) $W^{2}(y)$ is the conformal factor and $d\mu(y)$ is the
invariant volume (we have integrated out all the dependence on
other coordinates except $y=\cos^{2}\xi-\sin^{2}\xi$). This leads
to the explicit form of the orthogonality condition in the new 
variable:
\begin{equation}
\frac{C^{2}_{1}9^{3}(n)!(m)!(\alpha)!^{2}}{2^{\alpha+\beta+
\frac{11}{6}}(n+\alpha)!(m+\alpha)!}\int(1-y)^{\alpha}
(1+y)^{\beta}P^{(\alpha,\beta)}_{n}(y)P^{(\alpha,\beta)}_{m}(y)
dy=\delta_{nm}\ \label{1.32}
\end{equation}
By using the standard normalization condition for the Jacobi
polynomials:
\begin{equation}
\int(1-y)^{\alpha}(1+y)^{\beta}[P^{(\alpha,\beta)}_{n}(y)]^{2}dy=
\frac{2^{\alpha+\beta+1}}{2n+\alpha+\beta+1}
\frac{\Gamma(n+\alpha+1)\Gamma(n+\beta+1)}{(n)!
\Gamma(n+\alpha+\beta+1)} \label{1.33}
\end{equation}
one can determine $C_{1}$:
\begin{equation}
C_{1}^{2}=\frac{2^{\frac{5}{6}}}{9^{3}}(2n+\alpha+\beta+1)
\frac{\Gamma(n+\alpha+\beta+1)}{\Gamma(n+\beta+1)
\Gamma(\alpha+1)}\frac{\Gamma(n+\alpha+1)}
{\Gamma(n+1)\Gamma(\alpha+1)}
\label{1.34}
\end{equation}
Using that $J=\frac{\sqrt{25+16E^{2}}}{6}-\frac{19}{6}$, the
expression for the energy reads off
\begin{equation}
E^{2}=(J+\frac{7}{3})(J+4).
\label{1.35}
\end{equation}
We can now find the corrections to this energy by using the
semiclassical approximation:
\begin{equation}
\delta E^{2}=\frac{C^{2}_{1}9^{3}(n!)^{2}(\alpha!)^{2}}
{2^{\alpha+\beta+\frac{11}{6}}[(n+\alpha)!]^{2}}m^{2}
\lambda\int_{-1}^{1}(1-y)^{\alpha+1}
(1+y)^{\beta-1/3}[P^{(\alpha,\beta)}_{n}(y)]^{2}dy \label{1.36}
\end{equation}
Let us first calculate the integral:
\begin{equation}
I=\int_{-1}^{1}(1-y)^{\alpha+1}(1+y)^{\beta-1/3}[P^{(\alpha,\beta)}_{n}(y)]^{2}dy
\label{1.37}
\end{equation}
It is useful to  change variables by introducing
$y=1-\frac{x^{2}}{2n^{2}}$ after which
our integral becomes:
\begin{equation}
I=\int_{0}^{2n}\frac{x^{2\alpha+3}}{2^{\alpha+1}n^{4}}
(2-\frac{x^{2}}{2n^{2}})^{\beta-1/3}[\frac{1}{n^{\alpha}}
P^{(\alpha,\beta)}_{n}(1-\frac{x^{2}}{2n^{2}})]^{2}dx
\label{1.38}
\end{equation}
Using the well known asymptotic behavior of the Jacobi
polynomials for large $n$ we get:
\begin{equation}
[\frac{1}{n^{\alpha}}
P^{(\alpha,\beta)}_{n}(1-\frac{x^{2}}{2n^{2}})]
\approx(\frac{2}{x})^{\alpha}J_{\alpha}(x)
\label{1.39}
\end{equation}
where $J_{\alpha}$ are the Bessel functions of first kind. The
substitution of (\ref{1.39}) into the integral (\ref{1.38}) gives:
\begin{equation}
I=\int_{0}^{2n}\frac{x^{3}2^{\alpha-1}}{n^{4}}
(2-\frac{x^{2}}{2n^{2}})^{\beta-1/3}[J_{\alpha}(x)]^{2}dx
\label{1.40}
\end{equation}
The explicit solution is in terms of degenerate
hypergeometric function:
\begin{equation}
\begin{array}{l}
I=2^{\alpha+\beta+5/3}(\alpha+1)n^{2\alpha}
\frac{\Gamma(\beta+2/3)}{\Gamma(\alpha+\beta+8/3)
\Gamma(\alpha+1)}\times\\\\
_{2}F_{3}[\alpha+2,\alpha+1/2;\alpha+1,2\alpha+1,\alpha+\beta+8/3;-4n^2]
\label{1.41}
\end{array}
\end{equation}
For large $\alpha$ one can use the approximation:
\begin{equation}
\begin{array}{l}
_{2}F_{3}[\alpha+2,\alpha+\frac{1}{2};\alpha+1,2\alpha+1,
\alpha+\beta+\frac{8}{3};-4n^2]\rightarrow
\\\\\rightarrow\,_{2}F_{1}[\alpha+2,\alpha+\frac{1}{2};
\alpha+1;-\frac{-4n^{2}}{(2\alpha+1)
(\alpha+\beta+\frac{8}{3})}]
\label{1.42}
\end{array}
\end{equation}
To evaluate the correction to the energy (large $n$) we use the
transformation of the hypergeometric function:
\begin{equation}
\begin{array}{l}
_{2}F_{1}[a,b;c;z]=\frac{\Gamma(c)\Gamma(b-a)}
{\Gamma(b)\Gamma(c-a)}(-z)^{-a}\,_{2}F_{1}[a,1-c+a;1-b+a;\frac{1}{z}]+\\\\
\frac{\Gamma(c)\Gamma(a-b)}
{\Gamma(a)\Gamma(c-b)}(-z)^{-b}\,_{2}F_{1}[b,1-c+b;1-a+b;\frac{1}{z}].
\label{1.43}
\end{array}
\end{equation}
Evaluating the leading order behavior of the correction to the
energy, we find:
\begin{equation}
\delta E^{2}\approx(\frac{m^{2}\lambda}{2n}+(\alpha/2+1/4)
(2\alpha+1)(\alpha+\beta+8/3)\frac{m^2\lambda}{4n^{3}}).
\label{1.44}
\end{equation}

\subsection{Bohr-Sommerfeld quantization}
For completeness (and consistency check) we will give here also the
Bohr-Sommerfeld
quantization procedure applied to our case. First of all we point out
that our potential is obviously even, so
the quantized states satisfy:
\begin{equation}
(2n+\frac{1}{2})\pi=\int_{-\xi_{0}}^{\xi_{0}}\sqrt{E^{2}-m^{2}\lambda
\frac{\sin^{2}\xi}{(\cos\xi)^{2/3}}}d\xi
\label{1.45}
\end{equation}
In (\ref{1.45}) the values $\pm\, \xi_{0}$ are the turning points of the
potential, which are the solutions to the equation:
\begin{equation}
E=m\sqrt{\lambda} \frac{\sin\xi_{0}}{(\cos\xi_{0})^{1/3}}
\label{1.46}
\end{equation}
For large values of the energy $E$ we find that
$\xi_{0}=\pi/2$. It is appropriate to use the notation
$B=E/m\sqrt{\lambda}$, so if we now
define:
\begin{equation}
y=B^{-1}(\tan\xi)^{1/3}
\label{1.47}
\end{equation}
one can get for the measure $d\xi$:
\begin{equation}
d\xi=3B(\tan\xi)^{2/3}\cos^{2}\xi dy.
\label{1.48}
\end{equation}
With this at hand, we transform our integral to:
\begin{equation}
(2n+\frac{1}{2})\pi=3B^{3}m\sqrt{\lambda}\int_{0}^{y_{0}}
\sqrt{(B^{2}-\frac{y^{6}B^{6}}{(1+y^{6}B^{6})^{2/3}})
\frac{y^{4}}{(1+y^{6}B^{6})^{2}}}dy. \label{1.49}
\end{equation}
After some transformations and using the fact that
$B\rightarrow\infty$ we reduce our integral to:
\begin{equation}
(2n+\frac{1}{2})\pi=\frac{3}{B^{2}}m\sqrt{\lambda}
\int_{0}^{1}\frac{y^{2}\sqrt{1-y^{2}}}{y^{6}+ \frac{4}{3B^{6}}}dy.
\label{1.50}
\end{equation}
This integral is exactly solvable in terms of degenerate
hypergeometric functions $_{3}F_{2}(-\frac{4}{3B^{6}})$. Using the
series expansion for this function for small argument and taking into
account only the terms of order ${\cal O}(1/E)$ we end up with:
\begin{equation}
\frac{\sqrt{3}}{4}E-\frac{3^{1/6}4^{1/3}}{4}
\frac{m^{2}\lambda}{E}=2n+\frac{1}{2}.
\label{1.51}
\end{equation}
Solving this equation for $E$ and using that $E$ is very large we
find that the energy is:
\begin{equation}
E=\frac{4n+1}{\sqrt{3}}+\frac{3^{1/6}4^{1/3}}{2}\frac{m^{2}\lambda}{2n}
\label{1.52}
\end{equation}
So the classical energy, which through the correspondence is
identified with the bare dimension of the corresponding gauge operators,
is proportional to $n$:
\begin{equation}
\triangle_{b}=\frac{4n+1}{\sqrt{3}}
\label{1.53}
\end{equation}
According to the AdS/CFT correspondence, the anomalous dimension of
the corresponding SYM operator includes also the
corrections to the energy. In our case we find correspondingly
that:
\begin{equation}
\triangle-\triangle_{b}\approx\frac{m^{2}\lambda}{2n} \label{1.54}
\end{equation}

As expected, the result obtained by making use of Bohr-Sommerfeld
quantization is the same as the one calculated by using perturbation
theory. We can compare our solution to that of Minahan \cite{min}
for pulsating string on pure $S^{5}$. We see that the bare
dimension has the same behavior but the numerical coefficient is
different which is expected because our sphere $S^{4}$ is deformed
by the conformal factor. The anomalous dimension also has the same
behavior but with different numerical coefficient.

\sect{Conclusions}

In this section we summarize the results of our study.
The goal we pursued in this paper was to investigate the pulsating
strings in type IIA background described in section 3. The motivation
was the significant difference between the results found for
$AdS_5\times S^5$ background and those obtained in \cite{pope} for the
case of warped $AdS_6\times S^4$. We looked for the simplest pulsating string
solutions in warped $S^4$ part of $AdS_6\times S^4$ background. Using
the simple string ansatz 
\eq{
t=\tau \qquad \theta=m\sigma \qquad
\rho=\rho(\tau),
}
we find corresponding pulsating string solutions. 
Following the procedure developed in \cite{min}, we find the
energy. To obtain
the energy corrections we used the general approach suggested by
Minahan \cite{min}. For this purpose we consider the Nambu-Goto actions and find
the Hamiltonian. After that we quantize the resulting theory
semiclassically and obtain the corrections to the energy.
We calculate the corrections to the energy by making use
of two approaches, namely perturbation theory and Bohr-Sommerfeld
quantization. Both approaches give the same result
\eq{
\delta E^{2}\approx\lb[\frac{m^{2}\lambda}{2n}+(\alpha/2+1/4)
(2\alpha+1)(\alpha+\beta+8/3)\frac{m^2\lambda}{4n^{3}}\rb].
}
One can compare our solution to that of Minahan \cite{min}
for pulsating string on pure $S^{5}$. The comparison gives that the bare
dimension has the same behavior but the numerical coefficients are
different. This is expected since our sphere $S^{4}$ is deformed
by the conformal factor $W(\xi)$. From AdS/CFT point of view the 
corrections to the classical energy enter the anomalous 
dimensions of the operators in SYM theory and therefore 
they are of primary interest.
Therefore the anomalous dimensions
also have qualitatively the same behavior as in $AdS_5\times S^5$
but with different numerical coefficients.

As a final comment we note that to complete the analysis from AdS/CFT
point of view, it is of great interest to develop an
analysis allowing to compare our result to that in SYM side. We leave this
important question for future research.

\vspace*{.8cm}

{{\large{\bf Acknowledgements:}} R.R. would like to thank C. Nunez
  for comments and suggestions.

\end{document}